\documentclass[twoside,twocolumn,english,pra,aps,superscriptaddress,showpacs]{revtex4}
\usepackage[T1]{fontenc}
\usepackage[latin9]{inputenc}
\usepackage{color}
\usepackage{babel}
\usepackage{textcomp}
\usepackage{amsmath}
\usepackage{mathptmx}
\usepackage{amssymb,amsfonts,mathrsfs}
\usepackage{amsthm}
\usepackage{graphicx}
\usepackage[unicode=true,pdfusetitle,
bookmarks=true,bookmarksnumbered=false,bookmarksopen=false,
breaklinks=false,pdfborder={0 0 0},backref=false,colorlinks=true]
{hyperref}

\usepackage{breakurl}
\usepackage{epstopdf}
\usepackage{hyperref}

\makeatother
\theoremstyle{plain}

\hypersetup{
    colorlinks=ture,
    linkcolor=red
}

\makeatother
\theoremstyle{plain}

\makeatother

\begin{document}
	
\preprint{This line only printed with preprint option}

\title{The multiple re-entrant localization in a phase-shift quasiperiodic chain}

\author{Shan-Zhong Li}
\affiliation{Guangdong Provincial Key Laboratory of Quantum Engineering and Quantum Materials, School of Physics and Telecommunication Engineering, South China Normal University, Guangzhou 510006, China}

\affiliation{Guangdong-Hong Kong Joint Laboratory of Quantum Matter, Frontier Research Institute for Physics, South China Normal University, Guangzhou 510006, China}

\author{Zhi Li}
\email{lizphys@m.scnu.edu.cn}

\affiliation{Guangdong Provincial Key Laboratory of Quantum Engineering and Quantum Materials, School of Physics and Telecommunication Engineering, South China Normal University, Guangzhou 510006, China}

\affiliation{Guangdong-Hong Kong Joint Laboratory of Quantum Matter, Frontier Research Institute for Physics, South China Normal University, Guangzhou 510006, China}




\date{\today}

\begin{abstract}
Inspired by the recently discovered phenomenon of re-entrant localization (REL) [Roy et al., PRL 126, 106803 (2021)], we propose a new approach to induce REL, i.e., to control the quasiperiodic potential's phase-shift between odd and even sites, as thus the system can be dubbed as a phase-shift AAH model. We then analyze the participation ratios and corresponding scaling behaviors, and the results reveal that multiple re-entrant localization (MREL) phenomenon occurs. Furthermore, by depicting the behavior of extension dynamics, we obtain a whole visualized process of the system entering and re-entering the localized phase multiple times. Finally, we exhibit the distribution of quasiperiodic potential with different phase-shift and quasiperiodic parameter, and show the reason for the occurrence of MREL phenomenon, i.e., the introduction of phase-shift enables a part of eigenstates to escape from the localized phase, thus weakening the ``localizibility'' of the system.

\end{abstract}

\maketitle

\section{Introduction}
In 1958, by analyzing data from G. Feher and E. A. Gere's spin resonance experiment in polycrystalline silicon~\cite{GFeher1959a,GFeher1959b}, P. W. Anderson constructed a model of electrons with random disorder, where the quantum tunneling can be greatly subdued, thus leaving the electrons, which are originally free to move around in ordered systems, to be constrained~\cite{PWAnderson1958}. This phenomenon is known as Anderson localization. Previous studies have shown that in one-dimensional (1D) or 2D disordered systems, arbitrary disorder (even very weak) can nudge all the eigenstates of the system to enter the localized phase. However, for the 3D case, relatively weak disorder will give rise to mobility edges, which indicates the coexistence of localized and extended states~\cite{EAbrahams1979,PALee1985,FEvers2008,ALagendijk2009}. Since they can cause special thermoelectric response that has a number of potential applications in new thermoelectric devices, mobility edges are recently attracting a growing interest~\cite{RWhitney2014,KYamamoto2017,CChiaracane2020}.


Compared with the disordered system, the advantage of quasiperiodic systems lies in the clear critical point of extended-localized phase transition in low dimensional cases. That explains why quasiperiodic models are so frequently used in studying Anderson localization in 1D and 2D cases, among which Aubry-Andr\'e-Harper (AAH) model stands out as the most famous one for its self-duality:~all eigenstates of AAH model exhibit extended (localized) characteristics before (after) the quasiperiodic strength reaches a critical value~\cite{PGHarper1955,SAubry1980,MGoncalves2022}. In other words, the standard AAH chain has a precise critical point of extended-localized phase transition, and there is no intermediate phase, which means the extended and the localized states cannot coexist. Recently, relevant studies suggest that mobility edges can be induced by introducing long-range hopping~\cite{XDeng2019,NRoy2021} or modified quasiperiodic potential~\cite{SDSarma1988,SGaneshan2015,XLi2017,YWang2020,XLi2020,DDwiputra2022,HYao2019} into the AAH model, and precisely calculated with Avila theorem~\cite{YWang2020,DDwiputra2022,YWang2021a,YLiu2021a,YLiu2021b,YLiu2021c,YCZhang2022,ZHWang2022,ZHXu2022,YWang2023,XCZhou2022}. It is worth noting that in addition to exploring Anderson localization, AAH model can also be used to investigate various other phenomena, such as Hofstadter's butterfly~\cite{DHofstadter1976,CRDean2013}, many-body localization~\cite{SIyer2013,VMastropietro2015,SZhang2018,YYoo2020,DWZhang2020,YWang2021b,YfWang2022} and topological properties~\cite{FMei2012,YEKraus2012a,YEKraus2012b,LJLang2012,XCai2013,SGaneshan2013,MNChen2016,QBZeng2020,LZTang2021,LZTang2022}, etc.~\cite{SZLi2022}, which are all popular topics in research. Experimentally, AAH models have been realized in photonic crystal ~\cite{YLahini2009,YEKraus2012,MVerbin2013,MVerbin2015,PWang2020}, optical waveguide arrays~\cite{SAGredeskul1989,DNChristodoulides2003,TPertsch2004}, ultracold atomic~\cite{GRoati2008,GModugno2010,MLohse2016,SNakajima2016,HPLuschen2018,FAAnK2021}, polariton condensates~\cite{DTanese2014,VGoblot2020} and other experimental platforms~\cite{PRoushan2017,FAAn2018}. All these stand testimony to the significant value of AAH model in both theory and experiment.

On the other hand, re-entrant localiztion (REL) phenomenon has been recently found in the interpolating Aubry-Andr\'e-Fibonacci model~\cite{VGoblot2020,LJZhai2021,Atrkalj2021}. It is generally believed that after the extended-localized phase transition, the system will stay in the localized phase and remain unchanged with increasing disorder (or quasiperiodic) strength. However, the discovery of REL phenomenon challenged the traditional knowledge by revealing that, with the continuous enhancement of disorder (or quasiperiodic) strength, some eigenstates in REL systems would, after entering into the localized phase, ``jump out'' of it~\cite{VGoblot2020,SRoy2021,XPJiang2021,CWu2021,ZWZuo2022,APadhan2022,WHan2022,LZhou2022,RQi2023,ZLu2023}. In addition, recent work even uncovers a recurrent extension phase transition in the {\it p}-wave paired superconducting AAH chain~\cite{SZLi2023}. So far, the above novel phenomena have opened up a new avenue for grasping an overall picture of Anderson localization theory, and rightfully so, attracted great attention in both theory and experiment. We will propose in this paper, from a new perspective, an experimentally more available scheme to reveal the physical mechanism behind REL phenomenon. In concrete terms, we propose a phase-shift AAH model to investigate the corresponding localization properties, mobility edges and intermediate phases.

The rest of this article is structured as follows. We introduce the model in Sec.~\ref{Sec.2}. In Sec.~\ref{Sec.3}, we illustrate the multiple re-entrant localization phenomenon by participation ratios, scaling behavior, expansion dynamics and the a discussion of the effect of phase-shift on quasiperiodic structures. In Sec.~\ref{Sec.4},  we discuss the REL in the case of quasiperiodic potential with different irrational numbers. Main findings of this paper are concluded in Sec.~\ref{Sec.5}.

\section{The phase-shift quasiperiodic model}\label{Sec.2}
We start to prepare the model  by introducing a phase-shift $\Delta$ to even sites of AAH model, whose Hamiltonian reads
\begin{equation}
\label{Hami}
H=-J\sum_{j=1}^{L-1}(c_{j}^{\dagger}c_{j+1}+H.c.)+\sum_{j=1}^{L}V_{j}c_{j}^{\dagger}c_{j}
\end{equation}
with

\begin{equation}
V_{j}=
\lambda \cos[2\pi\alpha j+\Delta\frac{1+(-1)^{j}}{2}+\theta],
\end{equation}
where $c_{j}$ ($c_{j}^{\dagger}$) is the fermionic annihilation (generation) operator. $J$ and $L$ denote the strength of the nearest neighboring hopping and the system size. The parameters $\lambda$, $\theta$ and $\Delta$ correspond to the quasiperiodic potential's strength, global phase and phase-shift, respectively. One can see that when $\Delta=0$, the model reduces to the standard AAH model with the critical point of phase transition at $\lambda/J=2$~\cite{SAubry1980,MGoncalves2022}. Since we consider large system sizes, the effect of $\theta$ can be neglected~\cite{MGoncalves2022,APadhan2022}. Hereafter, unless otherwise specified, we set $\theta=0$, $\Delta \in \left[0,\pi \right]$, $\alpha=(\sqrt{5}-1)/2$. We use open boundary conditions in numerical calculation and $J=1$ as the energy unit.

\section{Multiple Re-entrant Localization}\label{Sec.3}
\subsection{participation ratio and scaling behavior}
Inverse participation ratios (IPR) and normalized participation ratios (NPR) are crucial quantities used to study localiztion, which read
\begin{equation}
\begin{split}
\xi_{m}=&\sum_{j=1}^{L}\left | \psi_{m,j}\right|^4,\\
\zeta_{m}=&\left(L\sum_{j=1}^{L}\left |\psi_{m,j}\right|^4\right)^{-1},
\end{split}
\end{equation}
where $\psi_{m,j}$ is the amplitude of the $m$th eigenstate on site $j$. Under the condition of $L\rightarrow\infty$, the extended state (localized state) corresponds to $\xi_{m}=0$ ($>0$) and $\zeta_{m}>0$ ($=0$). To better identify intermediate phases, one can define, based on the average IPR $\overline{\xi}=\frac{1}{L}\sum_{m=1}^{L}\xi_{m}$ and NPR $\overline{\zeta}=\frac{1}{L}\sum_{m=1}^{L}\zeta_{m}$, a physical quantity $\eta$ to depict such phases, which takes the form~\cite{SRoy2021,XLi2017,APadhan2022}
\begin{equation}
\eta=\log_{10}\left[\overline{\xi}\times\overline{\zeta}\right].
\end{equation}
This is because in intermediate phases, the extended state and the localized state can coexist. At this point the average IPR and NPR are both of finite values, so their products are not zero. Conversely, when $\eta$ approaches $0$, the system can be deduced as in a pure extended or localized phase. To be specific, the extended state and the localized state correspond to $\overline{\xi}\sim1/L$, $\overline{\zeta}\sim O(1)$ and $\overline{\xi}\sim O(1)$, $\overline{\zeta}\sim 1/L$, respectively. Since the system size $L>10^3$ in our calculation, the system will exhibit pure (intermediate) phase when $\eta<-3$ ($-3<\eta<-1$). 

 \begin{figure}[tbp]
	\centering	\includegraphics[width=8.7cm]{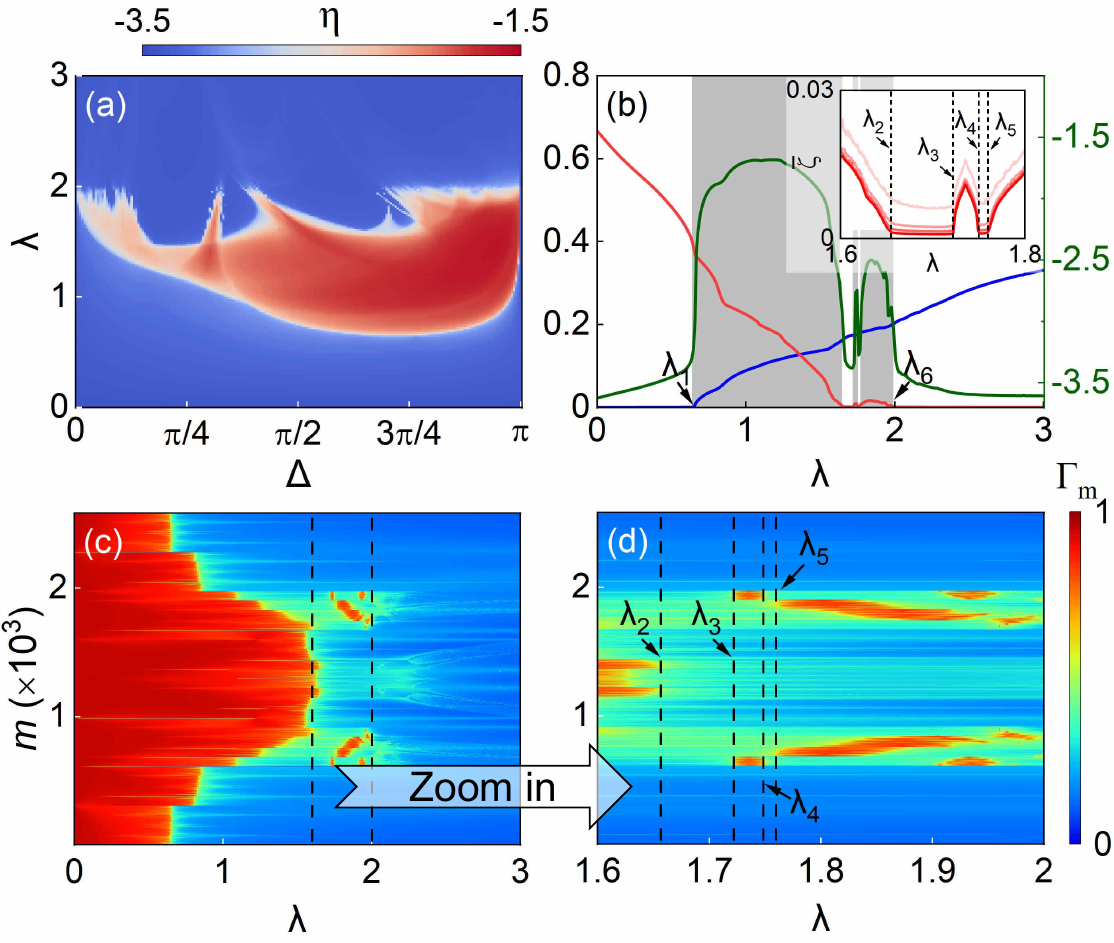}
	\caption{(a) $\Delta-\lambda$ phase diagram for system size $L=2584$. (b) IPR $\overline{\xi}$ (blue line), NPR $\overline{\zeta}$ (red line), and $\eta$ (green line) versus $\lambda$ for all eigenstates when $\Delta=3\pi/4$ and $L=4181$. The insert shows $\overline{\zeta}$ with $\lambda\in \left[1.6,~1.8\right]$, where $\lambda_{1,2,3,4,5,6}\approx 0.65,~1.655,~1.72,~1.75,~1.76,~1.99$ and system size $L=2584,~4181,~10946,~17711$ (from light to dark red), respectively. (c) Fractal dimension $\Gamma_m$  versus $\lambda$ at $\Delta=3\pi/4$. (d) Amplification of the region $\lambda\in\left[1.6,~2\right]$ of (c). We set the system dimension $L=2584$ in both (c) and (d).}\label{PD}
\end{figure}

In Fig.~\ref{PD}(a), we provide the $\Delta-\lambda$ phase diagram. It can be seen that REL will emerge in the system as the parameter $\lambda$ or $\Delta$ changes. Besides, we plot the variation of average IPR (left axis), average NPR (left axis), and $\eta$ (right axis) versus $\lambda$ under the fixed $\Delta=3\pi/4$ in Fig.~\ref{PD}(b), where we use grey areas to represent the intermediate phases of $\overline{\xi}>0$, $\overline{\zeta}>0$, $\eta>-3$ and the critical points $\lambda_1-\lambda_6$ are marked. As shown in the figure, when $\lambda>\lambda_{1}$, the system enters the intermediate phase. Then, by continuously increasing the quasiperiodic potential's strength, the system behaves as a pure localized phase when $\lambda_2<\lambda<\lambda_3$. After that, it will experience two similar REL processes. Finally, the whole chain will stay in localized phase, no matter how $\lambda$ increases. On closer inspection, we calculate the fractal dimension of the $m$th eigenstate, which is defined as
\begin{equation}
\Gamma_{m}=-\lim_{L\rightarrow \infty}\frac{\ln \xi_{m}}{\ln L}.
\end{equation}
Fractal dimension is an important indicator in the diagnosis of localization properties~\cite{YWang2020,YWang2022}. For the localized (extended) state, $\Gamma\rightarrow 0$ ($\rightarrow 1$), while the critical state corresponds to $\Gamma\in(0,1)$. We show in Fig.~\ref{PD}(c)(d) how the fractal dimension of different eigenstates changes with the parameter $\lambda$. One can see that in the system, the fractal dimension of some eigenstates $\Gamma\rightarrow 1$ when the quasiperiodic stength ranges $\lambda_3<\lambda<\lambda_4$ and $\lambda_5<\lambda<\lambda_6$, which indicates that a part of the eigenstates return to the extended state. This is due to the modification of the quasiperiodic potential by the phase-shift, in other words, the externally-imposed phase-shift makes the original quasiperiodic potential become the superposition of four sets of large periodic potential. As we know, the effect of large periodic potential is between that of periodic and quasiperiodic potential, thus leaving room for the emergence of novel mobility edge. This may explain the mechanism behind REL (or MREL) phenomenon here as well as in previous studies. Careful readers may notice that the change of NPR in the latter two REL processes is very slight [see Fig.~\ref{PD}(b)], which will be discussed in below.

We conduct scaling analysis on the indicators, so as to double-check the above finding. The effects of system size $L$ on $\eta$ and $\overline{\zeta}$ are calculated in parameter intervals of different phases, respectively. Specifically, as the size of the system increases, the $\eta$ corresponding to the extended phase or the local phase decays linearly, while the intermediate phase corresponds to a stable value. Although $\eta$ can effectively distinguish between the presence and the absent of a mobility edge, the scaling behavior of the average NPR $\overline{\zeta}$ must be studied in order to distinguish between the localized phase and the extendeded phase. As the  system size increases, the average NPR $\overline{\zeta}$ of the localized phase will decay linearly, while the intermediate phase and the extended phase will gradually converge to a stable value, respectively. Therefore, one can combine the scaling behavior of $\eta$ and $\overline{\zeta}$ to judge the phase of the system, that is, the corresponding $\eta$ of the extended phase exhibits linear decrease, while the average NPR $\overline{\zeta}$ is stable. Both $\eta$ and $\overline{\zeta}$ tend to stabilize eventually for the intermediate phase. As for the localized phase, both $\eta$ and $\overline{\zeta}$ exhibit linearly decrease.

 \begin{figure}[htbp]
	\centering
	\includegraphics[width=8.5cm]{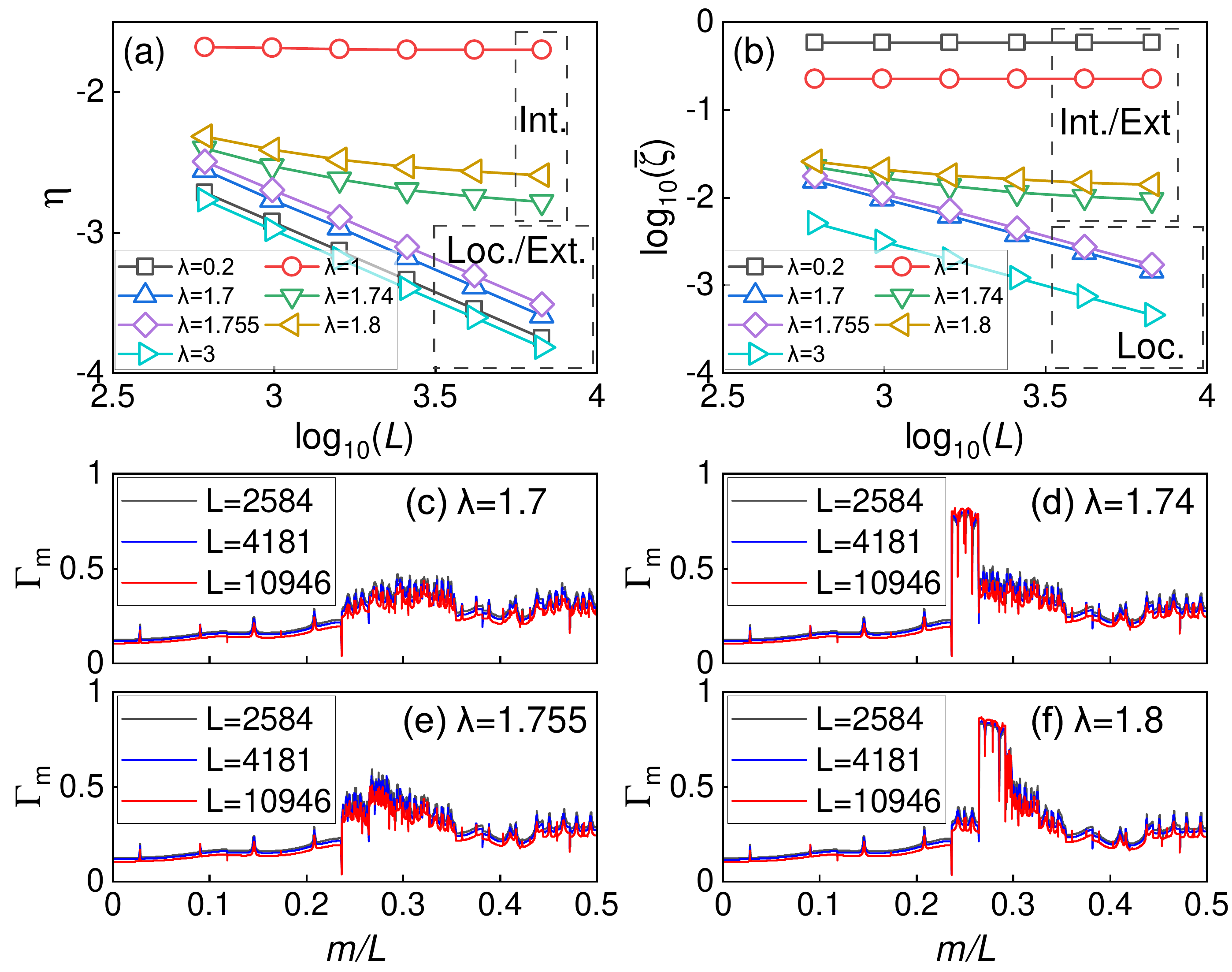}
	\caption{The scaling behavior of $\eta$ (a) and $\log_{10}(\overline{\zeta})$ (b) versus $\log_{10}(L)$ for different $\lambda$, where Ext., Loc., and Int. are abbreviations of extended, localized, and intermediate phases, respectively. (c-f) The behavior of $\Gamma_m$ with respect to the system size, where $\Delta=3\pi/4$ and other parameters are marked.  }\label{SA}
\end{figure}

From Fig.~\ref{SA}(a), one can find that for the intermediate phase ($\lambda=1,~1.74,~1.8$), the corresponding $\eta$ slowly tends towards a fixed value as the system size increases, while $\eta$ of the pure phase ($\lambda=0.2,~1.7,~1.7555,~3$) will linearly decrease with the system size. Besides, by observing the scaling behavior of the average NPR $\overline{\zeta}$, one can tell the extended and the localized pure phases apart. To be specific, as the system size grows, an approximation to a stable value of $\overline{\zeta}$ indicates the intermediate or the extended phase of the system; while for the case of localized phase ($\lambda=1.7,~1.7555,~3$), since $\overline{\zeta}\sim\frac{1}{L}$, $\overline{\zeta}$ will linearly decrease with the increasing $\log_{10}(L)$. Comparative analysis on the results of Fig.~\ref{SA}(a)(b) proves the existence of three different phases, i.e., the extended ($\lambda=0.2$), localized ($\lambda=1.7,~1.755,~3$) and intermediate phase ($\lambda=1,~1.74,~1.8$). 

Hereinbefore, we have seen that NPR changes slightly in the latter two REL processes [see Fig.~\ref{PD}(b)]. For accuracy, we calculate the scaling behavior of the corresponding fractal dimension. The results show that with the increase of quasiperiodic strength $\lambda$, the fractal dimension corresponding to a part of eigenstates transforms from localized ($\Gamma_m$ decreases with an increasing $L$) to extended ($\Gamma_m$ increases with an increasing $L$) properties, which causes the simultaneous emergence of localized and extended states in the system, thus giving rise to the intermediate phase as well as mobility edges [see Fig.~\ref{SA}(c)-(f)]. Since only a small number of eigenstates are transformed, the change of NPR is very small. The above results provide solid evidence that it is the introduction of phase-shift that weaken the localization property, hence resulting in the interesting MREL phenomenon.

\subsection{Expansion dynamics}

\begin{figure}[htbp]
	\centering
	\includegraphics[width=8.2cm]{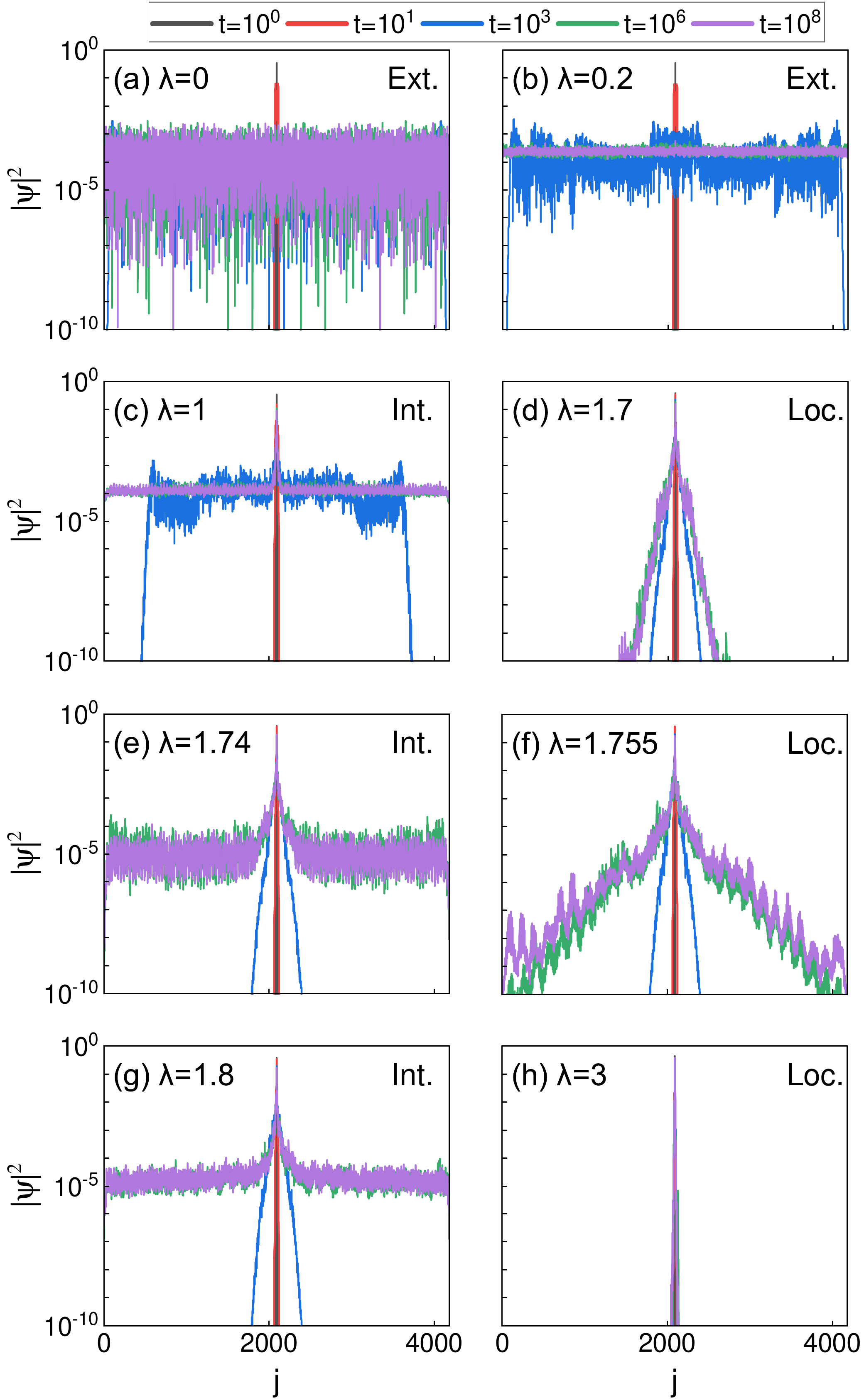}
	\caption{Real space probability distribution $\left|\psi_j(t)\right|^2$ for different $\lambda$ at $t=10^{0}$ (black), $10^{1}$ (red), $10^{3}$ (blue), $10^{6}$ (green), $10^{8}$ (purple), respectively. For all plots, we average $\left|\psi_j(t)\right|^2$ with different $\theta$ 100 times to smooth the data, throughout we set system size $L=4181$ and the parameter $\Delta=3\pi/4$. }\label{evo}
\end{figure}

In this section, we analyze MREL phenomenon from another perspective, i.e., expansion dynamics. Without loss of generality, we select the initial state $\left|\psi(t=0)\right>=\delta_{j,j_{0}}$, where $j_{0}=L/2$ is the initial position of wave function. By substituting Hamiltonian~\eqref{Hami} into the equation of motion, one can obtain the wave function versus time $t$, i.e., $\left|\psi(t)\right>=\exp(-iHt)\left|\psi(t=0)\right>$. Then we calculate the evolution of $\left|\psi(t)\right|^2$ with parameter $\lambda=0,~0.2,~1,~1.7,~1.74,~1.755,~1.8,~3$, respectively [see Fig.~\ref{evo}(a)-(f)]. 

The results reveal that when the system is in the extended phase [Fig.~\ref{evo}(a)(b)], the wave function will, after a long time of evolution, be evenly distributed in space; however, when the system is in the localized phase [Fig.~\ref{evo}(d)(f)(h)], the wave function will localize around the initial position, with its probability distribution decaying exponentially. For the case of intermediate phase, since both extended and localized states coexist in the system [Fig.~\ref{evo}(c)(e)(g)], the wave function will be partly localized, with the rest part evenly distributed. The results of expansion dynamics under different phase conditions well prove the emergence of MREL phenomenon. 

\begin{figure}[htbp]
	\centering
	\includegraphics[width=8.5cm]{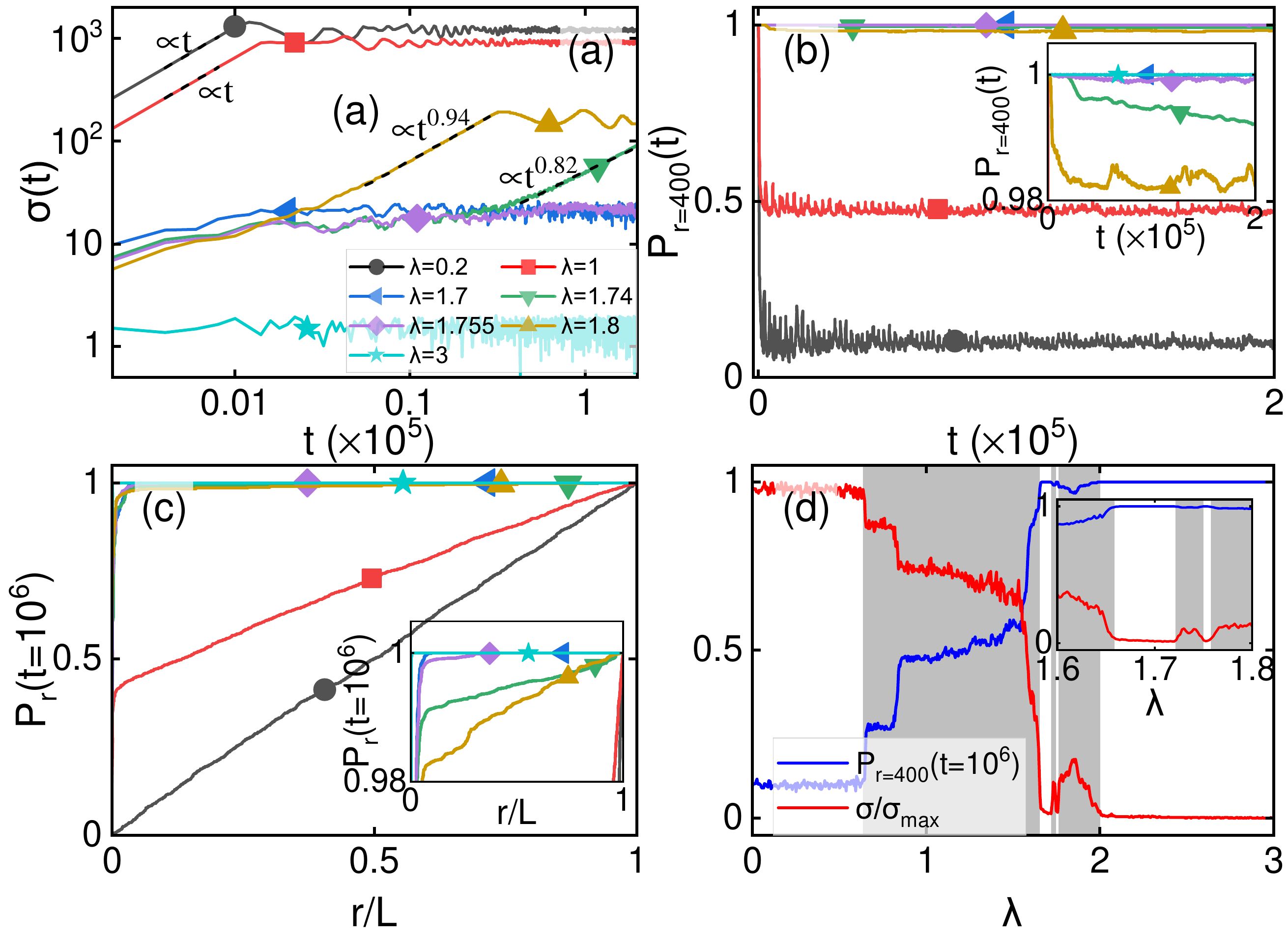}
	\caption{(a) $\sigma(t)$ versus $t$ for different $\lambda$ in log-log plane. The black dashed line indicates a power-law fit. (b) $P_{r=400}(t)$ versus $t$ for different $\lambda$. (c) The behavior of $P_{t=10^6}(r)$ varies with parameter $\lambda$. The inserts of (b) and (c) show the region of $P_{r=400}(t)\in[0.98,1]$ and the region of $P_{t=10^6}(r)\in[0.98,1]$, respectively. (d) $\sigma/\sigma_{max}$ (blue line) and $P_{r=400}$ (red solid) at $t=10^6$ for different $\lambda$, where $\sigma_{max}$ denotes the maximum of $\sigma(t=10^6)$ in the interval of $\lambda\in [0,3]$. Throughout, the system size $L=4181$ and $\Delta=3\pi/4$.}\label{diffuse}
\end{figure}

Furthermore, to better investigate the dynamical properties of wave function, one can usually resort to the root mean-square displacement~\cite{APadhan2022,ZJZhang2012,PBZeng2018,HPLuschen2018b,CMDai2018,ZXu2020}, which is defined as
\begin{equation}
\sigma(t)=\sqrt{\sum_{j}(j-j_{0})^2\left|\psi_{j}(t)\right|^2}.
\end{equation}
We calculate the root mean-square displacement and show $\sigma(t)$ with different $\lambda$ on the log-log plane in Fig.~\ref{diffuse}(a). The results reveal that $\sigma(t)$ corresponding to the extended phase ($\lambda=0.2$) grows rapidly in the early stage of evolution and then remains a fixed constant for a long time. The reason is that the wave function of the extended state spreads fast and can quickly be distributed evenly in space, hence $\sigma(t)$ is able to reach a large saturation value in a relatively short period of time. However, since the wave function of the localized phase ($\lambda=1.7,~1.755,~3$) almost does not spread, the corresponding $\sigma(t)$ is always small. The wave function of the intermediate phase ($\lambda=1,~1.74,~1.8$), in between of the above two, spreads partly and slowly outwards, therefore  $\sigma(t)$ grows relatively slowly, and the final saturation value also falls somewhere between the extended and the localized cases. By data fitting, one can find that the curves of extended phase ($\lambda=0.2,1$) and intermedia phase ($\lambda=1.74,1.8$) can be described by the expression $\sigma \propto t^{\beta}$ in the growth process, where $\beta\approx 1(\approx0.82, \approx0.94)$ for $\lambda=0.2$ or $1(=1.74,=1.8)$. The $\sigma(t)$ curves of localized phase maintains a stable value for a long time. This is because a wave packet in the expanded phase will quickly spread throughout the whole chain, so the final saturation value of $\sigma(t)$ coresponds the width of the whole chain. The wave function of localized phase will rapidly localized in a relatively small range, and therefore a relatively small stable value. The intermediate phase has both localized and extended properties, so the corresponding behavior of $\sigma(t)$ is in between.

In addition to the root mean-square displacement, survival probability is also a key physical quantity to study the localization properties of wave packet evolution~\cite{XDeng2019,APadhan2022,ZXu2020}, which reads
\begin{equation}
P_{r}(t)=\sum_{j_{0}-r/2}^{j_{0}+r/2}\left|\psi_{j}(t)\right|^{2},
\end{equation}
where $r$ denotes the interval of survival probability centered on $j_0$. By fixing the parameter $r$ ($t$), we calculate $P_r(t)$ versus $t$($r/L$) under different $\lambda$ [Fig.~\ref{diffuse}(b)(c)]. Let us first show the case of $r=400$ [see Fig.~\ref{diffuse}(b)]. Since the localized wave function ($\lambda=1.7,~1.755,~3$) is hardly diffusible and can remain within the range of $r$ for a long time, it features a relatively large value, i.e., $P_{r}(t)\rightarrow 1$. On the contrary, wave function of the extended phase ($\lambda= 0.2$) spreads rapidly over time along the entire chain, so the corresponding $P_{r}(t)$ decays fast to approach $0$. Similarly, for the intermediate phase, $P_{r}(t)$ can be reduced to a finite value due to its partial localization property.

On the other hand, since the final state can well reflect the physical properties of the system, we analyze the case of $t=10^6$ after a long period of evolution [see Fig.~\ref{diffuse}(c)]. It is not difficult to find that, $P_r(t)$ corresponding to the localized phase ($\lambda=1.7,~1.755,~3$) rapidly approaches $1$ as $r/L$ increases, while for the case of extended phase ($\lambda=0.2$), $P_r(t)$ tends towards $1$ almost linearly with the increasing $r/L$. The case of the intermediate phase, however, is quite interesting. Since the system is partially localized and partially extended, as $r/L$ increases, $P_r(t)$ will first shoot up to a value between $0$ and $1$, and then linearly grow to $1$ [see Fig.~\ref{diffuse}(c) and the insert].

Finally, we compare and analyze how the root mean-square displacement and survival probability evolve with the quasiperiodic strength $\lambda$ after a long time. As shown in Fig.~\ref{diffuse}(d), $\sigma(t)$ and $P_r(t)$ capture very well the key points in the ebb and flow of the system's localization property and exhibit MREL phenomenon, which are in good agreement with Fig.~\ref{PD}(b).

 \begin{figure}[thbp]
	\centering	\includegraphics[width=8.2cm]{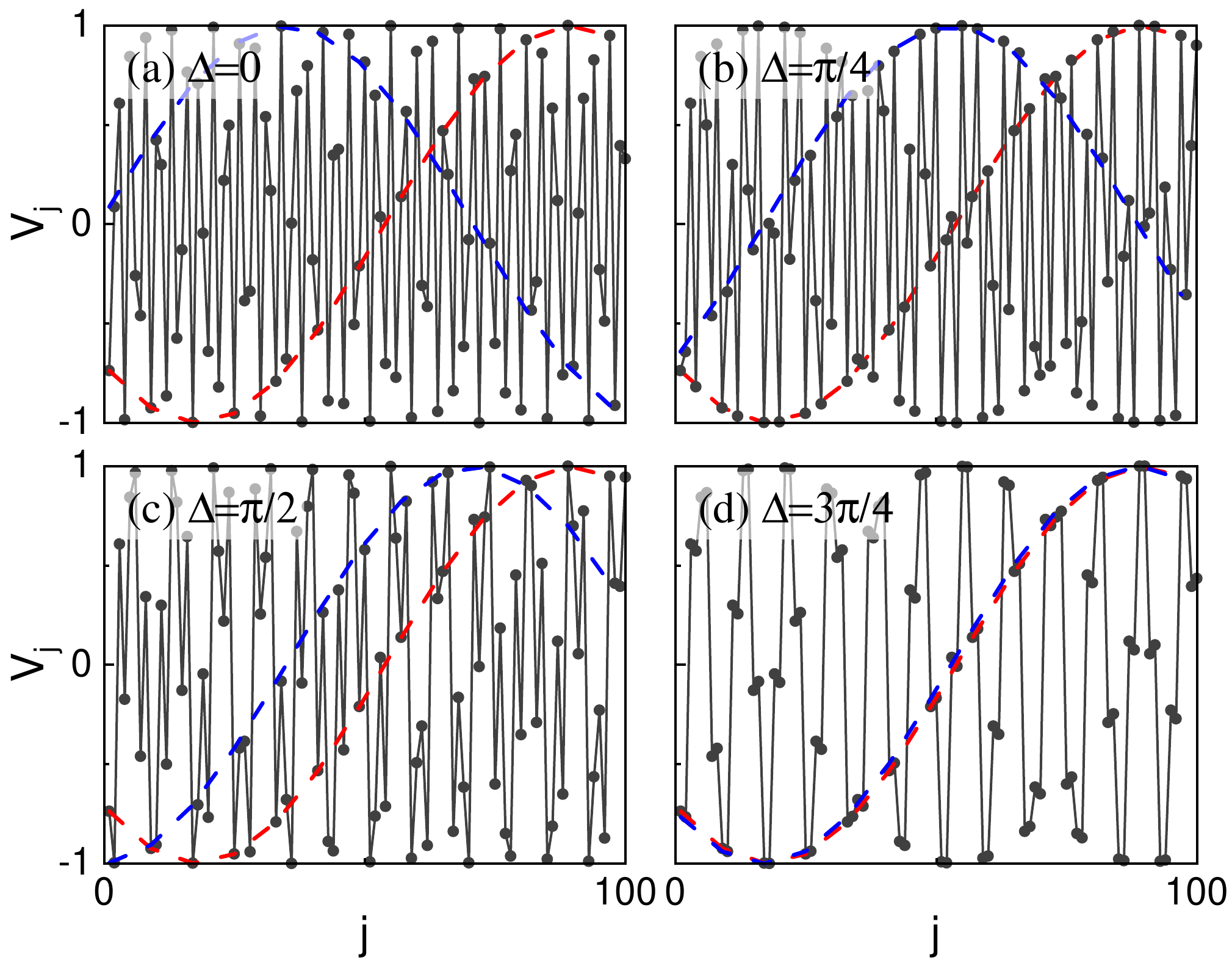}
 	\caption{(a) The quasiperiodic potential distribution of the phase-shift AAH model for (a) $\Delta=0$ (Standard AAH model), (b) $\Delta=\pi/4$, (c) $\Delta=\pi/2$, and (d) $\Delta=3\pi/4$. The other parameters $\lambda=1$, $\alpha=(\sqrt{5}-1)/2$. }\label{F5}
\end{figure}
 \begin{figure}[htbp]
	\centering	\includegraphics[width=8cm]{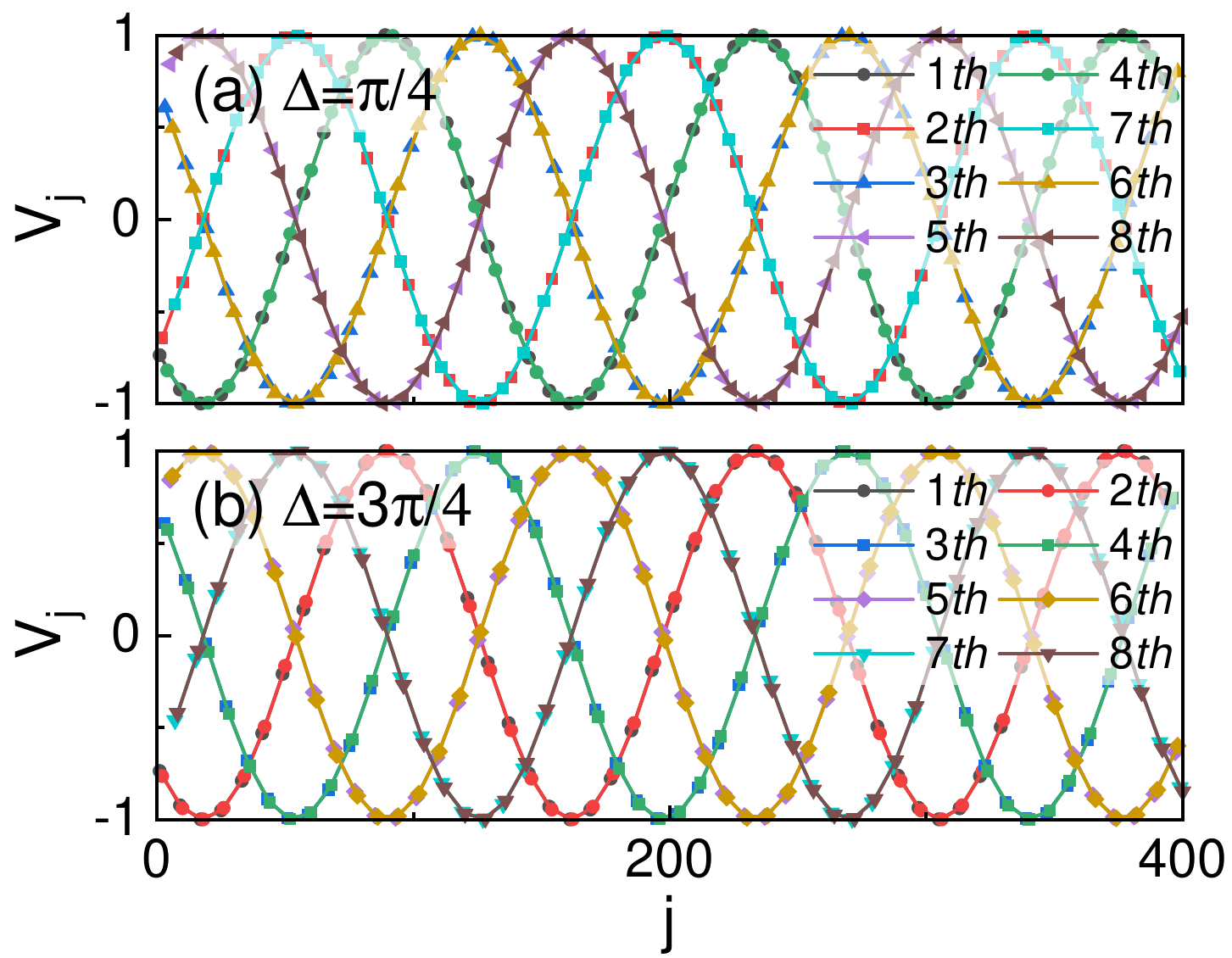}
	\caption{he quasiperiodic potential distribution after the introduction of a phase shift. (a) $\Delta=\pi/2$, at this point in the quasi-cell, the sublattices (1,4), (2,7), (3,6), (5,8) share a common set of quasiperiods. (b)$\Delta=3\pi/4$, in this quasi-cell, the neighbouring sublattices (1,2), (3,4), (5,6), (7,8) share a common set of quasiperiods. }\label{F6}
\end{figure}
\subsection{Phase shift induced changes in quasiperiodic potential structures}
The structure of the quasiperiodic potential is significantly modified by the introduction of a phase-shift. We present the variations in the potential structure for different $\Delta$ in Fig.~\ref{F5}(a)-(d), aiming to establish the possibility of inducing a REL potential structure. In the case of $\alpha=(\sqrt{5}-1)/2$, the AAH model's potential can be understood as an eight-sublattice unit cell, each with the same large period. We depict this large period composed of the first (red) and second (orange) sublattices with dashed lines. Remarkably, the introduction of a phase-shift leads to changes in the potential structure, gradually diminishing the quasiperiodic characteristics and restoring periodicity at specific phase-shift values (e.g.,Fig.~\ref{F5}(b)(d)). This resurgence of periodicity could be a crucial aspect in the emergence of REL in the phase-shifted AAH model.

 \begin{figure}[thbp]
	\centering
	\includegraphics[width=8.5cm]{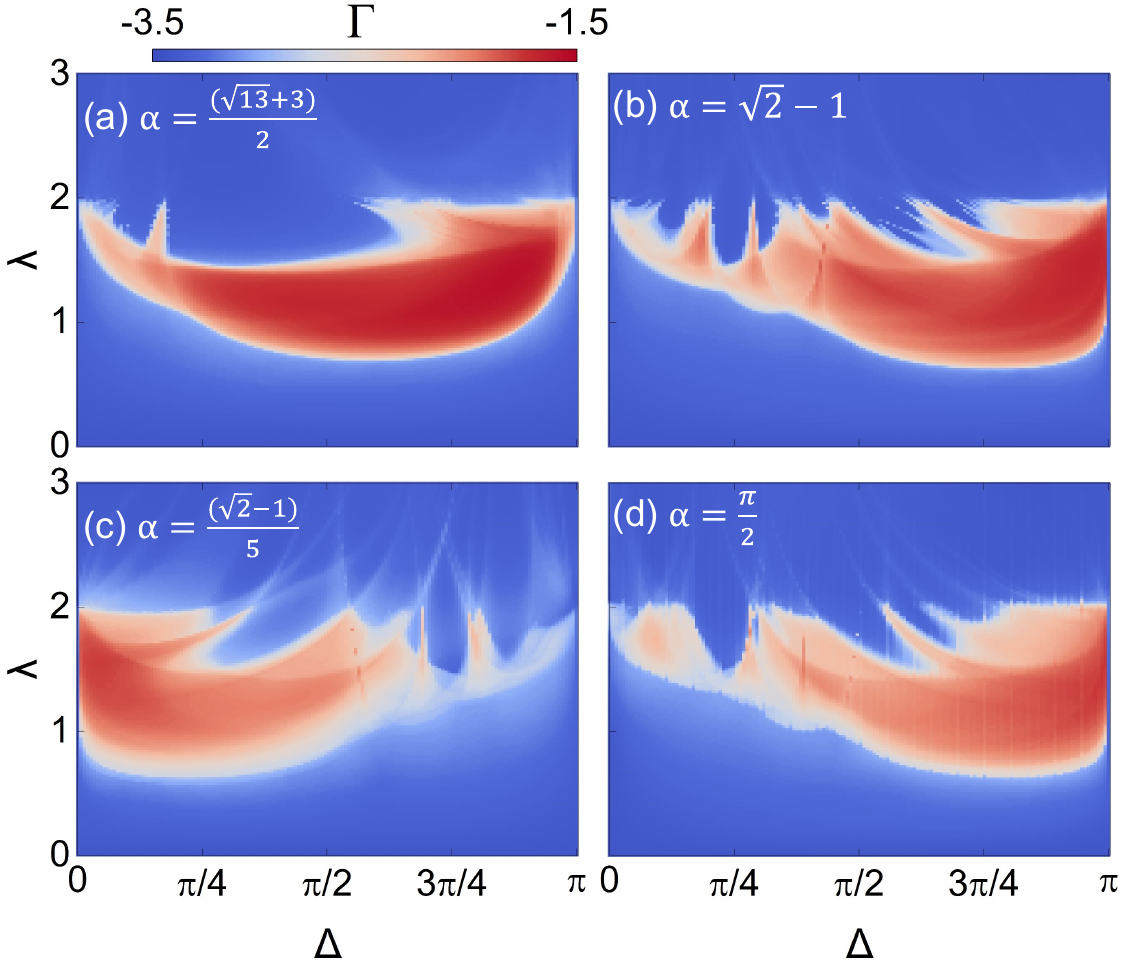}
	\caption{$\Delta-\lambda$ phase diagram for (a) $\alpha=(\sqrt{13}+3)/2$, (b) $\alpha=\sqrt{2}-1$, (c) $\alpha=(\sqrt{2}-1)/5$, and (d) $\alpha=\pi/2$. Throughout, the system size $L=2584$. }\label{F7}
\end{figure}
Additionally, we illustrate the formation of periods by all sublattices for two specific phase-shift values, $\Delta = \pi/4$ and $\Delta = 3\pi/4$, in Fig.\ref{F6}. As depicted in the figure, these phase-shift values lead to the emergence of two distinct types of sublattices with identical periodicity. For $\Delta = \pi/4$, the sublattices (1,4), (2,7), (3,6), and (4,8) share the same periodicity. This specific configuration, as shown in Fig.~\ref{PD}(a), corresponds to a regime in which the system exhibits a higher tendency for localization. Conversely, when $\Delta = 3\pi/4$ as shown in FIg.~\ref{PD}(b), the neighboring sublattices (1,2), (3,4), (5,6), and (7,8) exhibit the same periodicity. Under this condition, the system undergoes a manifestation of MREL. REL is possible when the potential periodicity of the system is restored and there is a common set of periodic potentials between the sublattice.


\section{Other quasiperiodic parameter}\label{Sec.4}
As discussed earlier, the quasiperiodic structure plays a crucial role in inducing mobility edges and REL. In addition, the effect of quasiperiodic parameters on REL is revealed in Ref.~\cite{PSNair2023}. In this section, we investigate the localization phase diagram for various quasiperiodic parameters $\alpha$.  We present the $\eta$ values in the $\lambda-\Delta$ plane at different $\alpha$ in Fig.~\ref{F7}(a)-(d). The results clearly demonstrate that different regions of REL exist for different $\alpha$, and the introduction of phase-shifts can initiate system localization.

For $\alpha=(\sqrt{13}+3)/2$ (the bronze ratio), REL occurs near $\Delta=3\pi/5$. Systems within the range of $\pi/4<\Delta<\pi/2$ are more likely to exhibit localization. When $\alpha=\sqrt{2}-1$ (the silver ratio), REL occurs in the vicinity of $\Delta=3\pi/5$. In the case of $\alpha=(\sqrt{2}-1)/5$, REL occurs around $\Delta=\pi/3$. For $\alpha=\pi/2$, REL is observed near $\Delta=2\pi/3$. It is worth noting that the introduction of phase-shifts can induce REL, and in some cases, even MREL.

\begin{figure}[tbp]
	\centering
	\includegraphics[width=8.5cm]{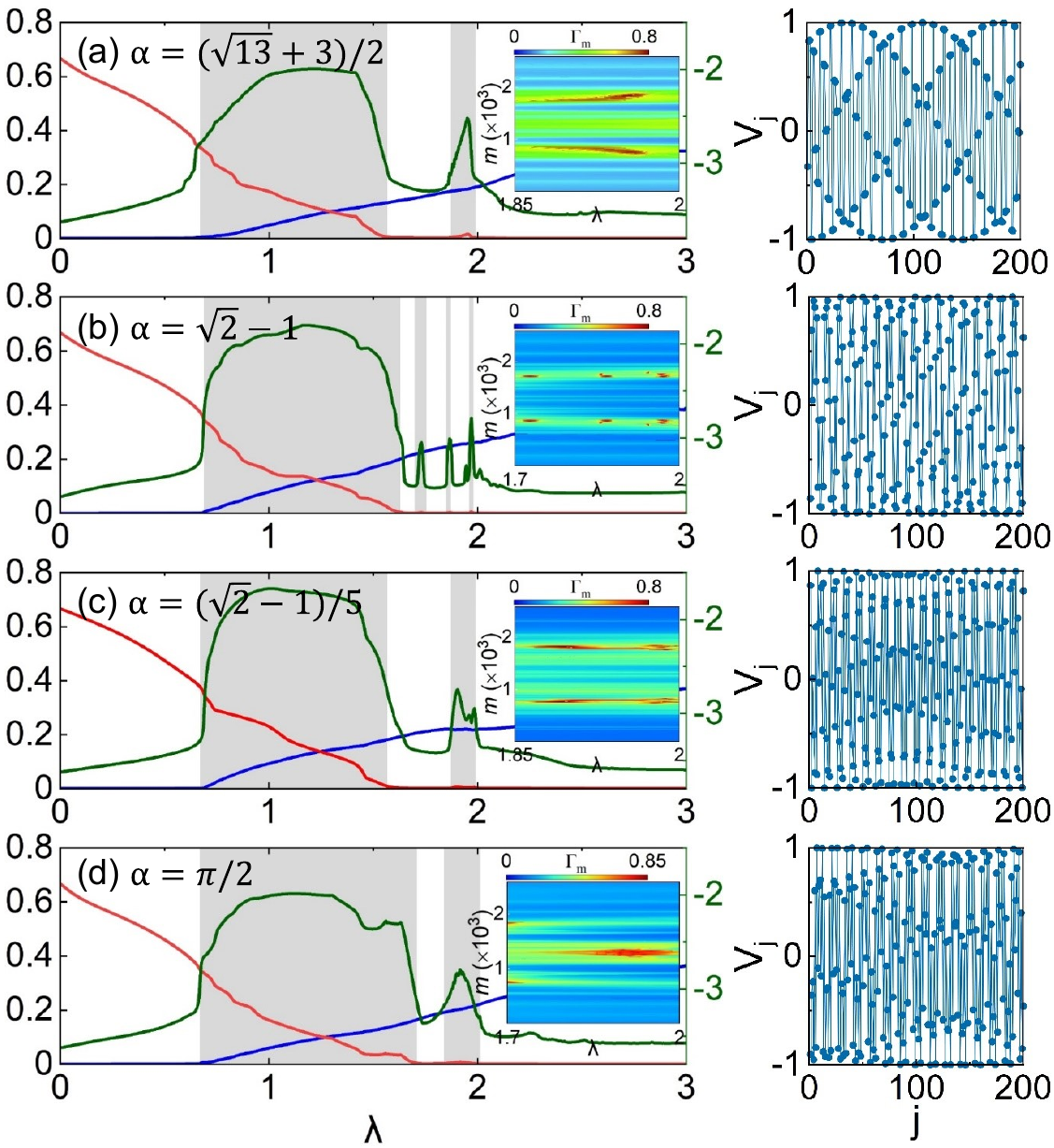}
	\caption{IPR $\overline{\xi}$ (blue line), NPR $\overline{\zeta}$ (red line), and $\eta$ (green line) versus $\lambda$ for (a) $\alpha=(\sqrt{13}+3)/2$, $\Delta=3\pi/5$, (b) $\alpha=\sqrt{2}-1$, $\Delta=3\pi/5$, (c) $\alpha=(\sqrt{2}-1)/5$, $\Delta=\pi/3$, and (d) $\alpha=\pi/2, \Delta=2\pi/3$. The inset shows the $\Gamma$ of all eigenstates in the REL region. The right column shows the quasiperiodic potential for the first two hundred lattices under the corresponding parameter, where $\lambda = 1$. For all main plots, the system size $L=4181$, and the system size $L=2584$ for inset.}\label{F8}
\end{figure}

We then provide a brief discussion on the region of REL by comparing the participation ratio $\overline{\xi}$ and $\eta$ in the left column of Fig.~\ref{F8}(a)-(d). It is evident that the increase in $\eta$ corresponds to the occurrence of REL, characterized by $\overline{\xi}$ being larger than zero again. The fractal dimension of all eigenstates within the REL region is shown in the inset of each figure, clearly illustrating the transition of some eigenstates from extended to localized phases as the quasiperiodic strength increases.

Furthermore, we present the quasiperiodic structure for the corresponding parameters in the right column. With the exception of $\alpha = \sqrt{2}-1$, the periodicity of the system's potential energy is restored. This phenomenon may explain why the phase-shifted AAH model can enhance localization, leading to the emergence of mobility edges and MREL.

\section{Conclusion}\label{Sec.5}

In summary, we propose a phase-shift AAH model and predict that MREL phenomenon can occur, i.e., the system can switch on and off its localization time and again. Besides, we prove the existence of mobility edge and intermediate phase in the system through approaches of participation ratios, scaling analysis and expansion dynamics. Furthermore, we have provided a possible explanation for the occurrence of REL, where the introduction of phase shifts suppresses the quasiperiodic structure, leading to a recovery of periodicity. Note that, the realization of REL and MREL requires harsh conditions in previous studies. To be specific, the introduction of complex quasiperiodic potentials~\cite{VGoblot2020} or the adjustment of lattice structure~\cite{SRoy2021} would be necessary based on the standard AAH model. The model proposed in this paper, however, can realize MREL phenomenon through phase-shift by readily manipulating the quasiperiodic potential, thus considerably lower the experimental requirements. Take AAH model in the ultracold atomic gases as an example. A phase-shift between the odd and even sites by controlling the laser is all what we need to construct the model, which is well achievable in terms of current technology~\cite{GRoati2008,GModugno2010,MLohse2016,SNakajima2016,HPLuschen2018,FAAnK2021}. We hope that the findings in this paper will bring benefit to the theoretical understanding of REL/MREL phenomenon as well as the future design of experimental platforms in related fields. 

\begin{acknowledgments}
We thank Shi-Liang Zhu, Xue-Jia Yu for helpful discussions and constructive suggestions. This work was supported by the National Key Research and Development Program of China (Grant No. 2022YFA1405300) and the Natural Science Foundation of Guangdong Province(Grants No.2021A1515012350).
\end{acknowledgments}




\end{document}